\documentclass{ws-ijmpa}
\usepackage[super, compress]{cite}
\usepackage{graphicx, amssymb, latexsym, epsfig, amsmath, bbm, color, hyperref,slashed,subfigure}
\hypersetup{colorlinks,urlcolor=black,citecolor=black,linkcolor=black,filecolor=black}
\usepackage{breakurl}

\newcommand{\Tr}{{\rm Tr}}

\newcommand{\cut}[1]{}

\newcommand{\psib}{\bar{\psi}}
\newcommand{\pa}{\partial}
\newcommand{\cQ}{{\cal Q}}
\newcommand{\cA}{{\cal A}}
\newcommand{\cAb}{\overline{\cal A}}
\newcommand{\cU}{{\cal U}}
\newcommand{\cUb}{\overline{\cal U}}
\newcommand{\cD}{{\cal D}}
\newcommand{\cDb}{\overline{\cal D}}
\newcommand{\cF}{{\cal F}}
\newcommand{\cFb}{\overline{\cal F}}
\newcommand{\KD}{{K\"{a}hler-Dirac }}
\newcommand{\cN}{{\cal N}}
\newcommand{\So}{\ensuremath{\mathrm{SO}}}
\newcommand{\Su}{\ensuremath{\mathrm{SU}}}
\newcommand{\imscale}{0.8}
\begin{document}
\markboth{G.~Bergner and S.~Catterall}{Supersymmetry on the lattice}

%
\catchline{}{}{}{}{}
%

\title{Supersymmetry on the lattice}

\author{Georg Bergner}
\address{Universit\"at Bern, Albert Einstein Center for Fundamental Physics, Institut f\"ur Theoretische Physik, Sidlerstr.~5, CH-3012 Bern, Switzerland\\
\href{mailto:g.bergner@uni-muenster.de}{g.bergner@uni-muenster.de}}

\author{Simon Catterall}

\address{Department of Physics, Syracuse University, Syracuse, NY 13244\\
\href{mailto:smcatter@syr.edu}{smcatter@syr.edu}}

\maketitle

\begin{history}
\received{Day Month Year}
\revised{Day Month Year}
\end{history}

\begin{abstract}
We discuss the motivations, difficulties and progress in the study of supersymmetric lattice gauge theories focusing in particular
on ${\cal N}=1$ and ${\cal N}=4$ super Yang-Mills in four dimensions. Brief reviews of the corresponding lattice formalisms are given and current results are
presented and discussed. We conclude with a summary of the main aspects of current work and prospects for the future.

\end{abstract}

\ccode{PACS numbers: 11.15.Ha, 11.30.Pb, 11.25.Tq, 12.38.Gc}

\section{Why study supersymmetric theories using lattice gauge theory?}

There are several reasons why supersymmetry has played such an important and prominent role in modern 
developments of quantum field theory. One is its versatility in constructing extensions of the Standard Model
of particle physics. Supersymmetry provides a solution to the hierarchy problem by providing a mechanism for
the cancellation of bosonic and
fermionic contributions to quantities such as the Higgs mass and as a bonus it includes natural candidates for dark matter. 
Furthermore, supersymmetric extensions 
of the standard model have implications at scales that are testable by current collider experiments. 
At the more fundamental level supersymmetry provides a bridge between the standard model and descriptions 
of quantum gravity based on supergravity and string theory. 

Supersymmetric extensions of the standard model like the minimal supersymmetric Standard Model (MSSM) are based on softly broken supersymmetric theories and the
relations to the experimental data are usually established by perturbative calculations. In general there are
very many of these soft breaking terms (approximately one hundred in the MSSM) and this yields to a lack of
predictability in such theories. It is generally believed that such softly broken theories should be thought of as
effective field theories describing the low energy behavior of a theory in which dynamical supersymmetry breaking has taken
place at some high scale and in some hidden sector. Presumably this breaking arises as a consequence of some
presently poorly understood strong dynamics. In principle lattice simulations of such strongly coupled supersymmetric theories
can allow one to measure such soft parameters in terms of a handful of non-perturbative quantities in a manner similar to the way
lattice QCD allows for the prediction of the low energy constants of chiral effective theory. En route to this ambitious goal
one must first understand the supersymmetric analog of the pure glue sector of QCD -- ${\cal N}=1$ super Yang-Mills theory (SYM). Numerical
studies of this theory form a major focus of this review.
The extension of this theory to include fermions in the fundamental representation of the
gauge group - super QCD - is currently too difficult for direct simulation in four dimensions. However we do discuss numerical
work that has been done in two dimensional super QCD. 

A second motivation for the study of supersymmetric theories is that they arise rather naturally
in systems including gravity in particular string theory. Of specific interest in this regard are
the holographic dualities which link the solution of classical gravitational systems with the strongly coupled behavior
of planar supersymmetric gauge theories. ${\cal N}=4$ super Yang-Mills theory furnishes the first and best understood
example of such a correspondence and numerical studies of this theory using novel lattice actions which preserve an element
of supersymmetry form a second major strand of this review.

Supersymmetry allows a better analytic understanding of quantum effects and offers, therefore, an interesting new perspective for
the investigations of strongly coupled theories. An example are the exact predictions\cite{Shifman:1987ia,Novikov:1983uc}, 
extensive semiclassical analysis\cite{Poppitz:2012sw}, and conjectured relations to QCD\cite{Armoni:2003fb} for ${\cal N}=1$ SYM. 
The verification and extension of the theoretical considerations is a further motivation for the 
numerical studies of supersymmetric theories. We elaborate a little bit more on this in the section about 
the results in ${\cal N}=1$ SYM. 

In the following we provide a short summary of four dimensional supersymmetric theories in the continuum to clarify the discussion.
The on-shell action of ${\cal N}=1$ SYM has the following form
\begin{equation}
 S_{\text{SYM}}=\int d^4 x\; \Tr \left[\frac{1}{4} F_{\mu\nu}^2+\frac{1}{2} \psib \slashed{D} \psi\right]\; ,
\end{equation}
where the field $\psi$ are Majorana fermions that transform in the adjoint representation of the gauge group ($D_\mu \psi=\pa_\mu \psi-ig[A_\mu,\psi]$).

The on-shell action of the ${\cal N}=1$ Wess-Zumino model contains a Majorana fermion $\psi$ and a complex bosonic field $\phi$,
\begin{equation}
 \label{eq:WZact}
 S_{\text{WZ}}=\int d^4 x\; \left[\partial_\mu \phi^\dag \partial^\mu \phi+|(m\phi+g\phi^2)|^2+\frac{1}{2}\psib(\slashed{\partial}+m+g\phi P_+ + g\phi^\dag P_- )\psi \right]\; ,
\end{equation}
where $W'(\phi)=m\phi+g\phi^2$ is the first derivative of the superpotential $W$ with respect to $\phi$ and $P_{\pm}=\frac{1}{2}(1\pm\gamma_5)$.  Supersymmetric QCD is a combination of ${\cal N}=1$ SYM, the pure gluonic sector, coupled to a matter sector
represented by an ${\cal N}=1$ Wess-Zumino model with fields in the fundamental representation.  ${\cal N}=2$ SYM corresponds to a variant of supersymmetric QCD without a 
superpotential and the matter fields in the adjoint representation.

${\cal N}=4$ SYM in four dimensions is derived from a dimensional reduction of ${\cal N}=1$ SYM in ten dimensions. The ten dimensional theory contains one Majorana-Weyl fermion that is reduced to 
four Majorana fermions and the six additional gauge field components become scalar fields in the adjoint representation. The action consists of $S_{\text{SYM}}$ with four fermion flavors,
Yukawa interactions with the six scalar fields $X_i$ that couple the different fermion flavors, and a bosonic action 
\begin{equation}
\label{eq:N4Bact}
 S_{\text{B}}=\int d^4 x\; \left[\frac{1}{2} D_\mu X^i D^\mu X^i+\frac{1}{4}[X^i,X^j]^2\right]\; .
\end{equation}

\section{Challenges with supersymmetric theories on the lattice}
The simulation of supersymmetric theories presents difficulties above and beyond those commonly encountered
in for example lattice QCD. These additional problems derive from the nature of supersymmetry itself as
the only non-trivial extension of the usual symmetries of spacetime. A consequence of this fact is that
the symmetry connects each fermion in the theory with a corresponding boson possessing the same quantum numbers. This
presents a problem since fermions and bosons are
usually handled differently in lattice simulations. Further considerations are related to common 
features of many supersymmetric theories: the bosonic potential in theories with extended supersymmetry naturally includes 
flat directions which must be regulated if they are to yield stable simulations. Additionally  the fermion representations
that commonly arise in these theories can lead to technical or even severe sign problems.

\subsection{Supersymmetry breaking by the lattice discretization}
Supersymmetry is connected with the symmetries of space time, which is commonly expressed in the terms of the simplified 
part of the supersymmetry algebra
\begin{equation}
 \{Q,Q\}\propto P_{\mu}\; ,
\end{equation}
where $Q$ are the generators of supersymmetry and $P_\mu$ the generators of translation. The lattice has no infinitesimal translation and, like the
symmetries of space-time, supersymmetry is broken by the lattice discretization. Typically, the remnant lattice symmetry ensures the restoration of 
the full space-time symmetry, but supersymmetry remains broken.

In a more detailed investigation one finds that the symmetry 
breaking is related to the violation of the Leibniz rule by any discrete derivative operator \cite{Dondi:1976tx,Kato:2008sp,Bergner:2009vg}.
The supersymmetry transformation of the action with the fermionic parameter $\epsilon$  generates terms of the form
\begin{equation}
 \epsilon \gamma^\mu\int d^4 x\; \left[ (\partial_\mu \phi_1)\phi_2 \psi +\phi_1 (\partial_\mu \phi_2)\psi+\phi_1 \phi_2 (\partial_\mu \psi)\right]=\int d^4 x\; \partial_\mu V^\mu\;,
\end{equation}
where the Leibniz (or chain) rule is applied to identify it with a total derivative.
The only way to maintain the Leibniz rule on the lattice and implement supersymmetry fully is by non-local derivative and product operators. 
Hence either locality or supersymmetry are 
violated on the lattice. 
In our review we apply the usual definition of locality in terms of an exponential decay of the operators 
with the distance on the lattice.
This resembles the Nielsen-Ninomiya theorem \cite{Nielsen:1981hk} concerning
the problem of constructing chiral lattice gauge theories. For chiral symmetry the Ginsparg-Wilson 
relation, a modified symmetry relation on the lattice, provides a practicable and general solution. 
In its general form it reads
\begin{equation}\label{eq:ggwrel}
 M^{ij} \phi_j \frac{\delta S}{\delta \phi_i}=(M \alpha^{-1})^{ij}\left(\frac{\delta S}{\delta \phi_j} \frac{\delta S}{\delta \phi_i}-\frac{\delta^2 S}{\delta \phi_j \delta \phi_i}\right)\; ,
\end{equation}
where the left hand side is the variation of the lattice action $S$ under the general transformation $\phi^i\rightarrow \phi^i+M^{ij}\phi_j$ and the 
right hand side corresponds to a controlled 
local 
breaking by a blocking kernel $\alpha$.
It reduces to a simple expression for chiral symmetry, but in the case of supersymmetry,
no practicable solutions have yet been found \cite{Bergner:2008ws}.
Note, furthermore, that the class of modified symmetries in \eqref{eq:ggwrel} is quite restricted due to the locality of the breaking term.

Since there is no generic solution, the representation of supersymmetry on the lattice is a model dependent issue. In general fine-tuning
ensures the restoration of the symmetry in the continuum limit. In models with extended supersymmetry, the implementation of a part of the symmetry
algebra ensures a significant reduction of this fine-tuning.

\subsection{Fermion doubling and fermion mass}
The fermion doubling problem is a well-known difficulty for the correct representation of fermionic fields on the lattice.
The background of this problem is the Nielsen-Ninomiya theorem, which states that
 locality and chiral symmetry can not be maintained without the introduction of additional fermionic degrees of freedom - the doublers. Furthermore the
 final lattice theory is then necessarily vectorlike.

Basically there are three options: one can employ non-trivial lattice fermion actions such as the lattice  K\"{a}hler-Dirac action  or (reduced) staggered
fermion action. This option is only open for certain theories which possess extended supersymmetry where the additional fermions that  arise in
these formulations can be interpreted as the correct number of continuum physical flavors.

Alternatively one can consider additional momentum dependent
mass terms, like the Wilson mass term, that remove the doubling modes in the continuum limit. These terms violate the equality between bosonic and fermionic masses unless they are introduced also in the bosonic sector \cite{Catterall:2000rv, Giedt:2004vb}. Bosons and fermions are then treated on the same footing with the same derivative operators and mass terms. The doubling problem is also introduced for the bosonic fields and, like for the fermions, the unphysical degrees of freedom are removed by additional mass terms. These mass terms can be consistently introduced in the superpotential corresponding to a modification of $m$ in \eqref{eq:WZact}. Hence not only the mass, but also higher vertices (proportional to $mg$ in \eqref{eq:WZact}) are modified in the on-shell formulation. The mass term breaks not only chiral symmetry but also corresponding bosonic symmetries. The application of this method for gauge fields, especially in a compact formulation, is not possible. 

A third option is to allow for nonlocal lattice actions. This might be a solution in lower dimensional theories without gauge fields. In the general case there is no proof of a well behaved continuum limit for a theory that violates locality. 

\subsection{Flat directions}
In theories with extended supersymmetry there are generically flat directions of the bosonic potential introduced by commutator terms of fields in the adjoint representation, see \eqref{eq:N4Bact}.
Such kinds of terms arise naturally in a dimensional reduction of pure Yang-Mills theories. In these pure bosonic cases the classical flat directions get usually lifted by quantum effects. Supersymmetry leads, however, to cancellations between the bosonic and fermionic contributions and the flat directions can survive in the quantum theory \cite{deWit:1988xki}. These effects are generically difficult to handle in numerical simulations - for example they can become unstable due to finite temperature
or lattice artifacts or the simulations may not be efficient at exploring the flat directions \cite{Catterall:2009xn}.
 One approach that has proven effective in theories with some exact supersymmetry (see later) is to modify the lattice action to include
additional scalar mass terms that lift the flat directions and to subsequently investigate the behavior of the observables when these regulator terms
are removed \cite{Catterall:2015ira}. 

\subsection{Sign problem in supersymmetric theories}
The Witten index measures the difference between bosonic and fermionic ground states 
in a supersymmetric theory.
If it is zero the theory can exhibit  spontaneous supersymmetry breaking.
 It is defined as
\begin{equation}
 \tilde{Z}=\Tr\left[ (-1)^F e^{-\beta H}\right]\; ,
\end{equation}
where, in contrast to the thermal partition function, $(-1)^F$ includes a minus sign for fermionic states of the Hamiltonian $H$.
It corresponds to a twisted partition function that is the sum of all differences between fermionic and bosonic energy states.
The usual thermal partition function 
employs periodic boundary conditions for the bosons and antiperiodic boundary
conditions for the fermions in one compact direction that corresponds to the temperature. The twisted partition function
has, instead, periodic boundary conditions for all fields.
If the Witten index is zero, there must necessarily
be the same number of negative and positive contributions from the configurations in the path integral with periodic boundary conditions. This means
a severe sign problem, or a zero by zero division in the computation of observables \cite{Catterall:2003ae}. A simple way to reduce, but not completely resolve, this sign problem is the application of supersymmetry breaking antiperiodic boundary conditions for fermion fields with a subsequent
extrapolation to the zero temperature continuum limit \cite{Wozar:2011gu}. More elaborate solutions are based on loop representations and Worm algorithms \cite{Steinhauer:2014yaa}.

Even in the case of a non-zero Witten index, a mild sign problem appears in several supersymmetric theories. In this case the negative contributions might be introduced by the discretization and disappear in the continuum limit.
One example is $\mathcal{N}=1$ supersymmetric Yang-Mills theory. Using Wilson fermions negative contributions from the Pfaffian are possible. They are enhanced towards the chiral limit, but suppressed in the continuum limit. In general it is relatively simple to handle these kind of sign problems, either by reweighting or by avoiding the 
critical parameter range. The only remaining challenge is the measurement of the Pfaffian sign.
\section{Solutions}
\subsection{Fine tuning}
\label{sec:finetuning}
The fact that typical lattice actions break supersymmetry leads to a proliferation of supersymmetry breaking counterterms in the effective action describing the
effects of quantum corrections. In general there are a large number of such relevant counterterms. To approach a supersymmetric continuum limit then requires that
all such terms be added to the bare lattice action and their coefficients carefully tuned as the lattice spacing is reduced. This is the famous fine tuning
problem of lattice supersymmetry.

This approach is particularly simple for super-renormalizable theories, where the coefficients can be calculated perturbatively. One of the earliest examples is the two dimensional Wess-Zumino model \cite{Golterman:1988ta}.
In supersymmetric Yang-Mills theories the possible counterterms are restricted by the gauge symmetry 
and the remnant space-time symmetry on the lattice. In $\mathcal{N}=1$ supersymmetric Yang-Mills theory
there is only one remaining counterterm: the gluino mass term \cite{Curci:1986sm}. It is the same tuning that is needed 
for the restoration of chiral symmetry in the continuum limit. The numerical determination of the coefficient is feasible and can be done using either the supersymmetric or the chiral Ward identities. If the fermion action fulfills the Ginsparg-Wilson relation, both chiral symmetry and supersymmetry
are ensured in the continuum limit.

The fine tuning is considerably more difficult in theories with scalar fields. These are part of the matter multiplet in supersymmetric QCD or appear
in the vector multiplet alongside the gauge field in theories with extended supersymmetry. If Ginsparg-Wilson fermions are employed, 
a combined tuning of several different counterterms needs  to be done \cite{Giedt:2009yd}. It might be guided by perturbative arguments as shown for the Wess-Zumino model \cite{Bartels:1982ue} and
for $\mathcal{N}=2$ supersymmetric Yang-Mills theory \cite{Montvay:1995rs}. 
In theories which preserve part of the supersymmetry algebra it is sometimes possible to reduce the number of fine tunings dramatically. A particular example of this
is $\mathcal{N}=4$ super Yang-Mills where a single tuning is all that is required to target the correct continuum theory.

\subsection{Preserving part of the supersymmetry algebra}
\label{sec:plsusy}
While discretization of supersymmetric theories generically breaks supersymmetry completely there are
situations where a subalgebra {\it can} be preserved. In many cases the existence of
this subalgebra places strong constraints on the possible counter term structure of
the theory and can reduce or even eliminate the fine tuning problem that
has been described earlier. 

These cases all involve theories
with extended supersymmetry - in fact in the case of
pure super Yang-Mills theories  the precise constraint is that the number of real fermionic degrees
of freedom must be $2^D$ where $D$ is the (Euclidean) spacetime dimension. 

\subsubsection{Two dimensional super Yang-Mills}
Let us see how this works in perhaps the simplest example: $(2,2)$ super
Yang-Mills in two dimensions \cite{Unsal:2006qp,Catterall:2007kn,Catterall:2009it}. The field content of this theory corresponds to two degenerate
flavors of Majorana fermions 
$\lambda^I,\,I=1, 2$,
two scalar fields $B^I$ and a gauge field $A_i,\,i=1,2$.
The global symmetries of the theory include $\So_{\rm Lorentz}(2)$ and a flavor
symmetry $\So_{\rm flavor}(2)$ and allow
one to decompose the fields of the theory under a {\it twisted} rotational symmetry corresponding to the
diagonal subgroup 
\begin{equation}
\So^\prime(2)={\rm Diag}\left(\So_{\rm Lorentz}(2)\times \So_{\rm flavor}(2)\right)\end{equation}
Under this twisted symmetry the fermions transform like a 2d matrix $\Lambda$
\begin{equation}
\lambda^I_\alpha\to G^{IJ}\lambda^J_\beta \left(G^T\right)_{\beta\alpha}\end{equation}
with $G$ a $\So(2)$ transformation.
Given this matrix structure it is then natural to expand $\Lambda$ on products of two dimensional gamma matrices
\begin{equation}
\Lambda=\eta I+\psi_i\sigma_i+\chi_{12}\sigma_1\sigma_2\end{equation}
The appearance of the scalar fermion $\eta$ is crucial - it implies the existence of a scalar supersymmetry
$\cQ$ and from the original supersymmetry algebra 
it is easy to show that $\cQ$ satisfies the subalgebra $\{\cQ,\cQ\}=0$. The absence of a
generator of translations on the RHS of this expression means that this supercharge can coexist
with a discrete lattice. Indeed it is possible to show that the action of the theory can be
written in a $\cQ$-exact form
\begin{equation}
S=\cQ \int d^2x\, \Tr\left( \chi_{12}\cF_{12}+\eta[\cDb_i,\cD_i]+\frac{1}{2}\eta d\right)\end{equation}
The gauge field $\cA$ entering in this expression is not the original gauge field but a complexified
field taking the form $\cA_i=A_i+iB_i$ containing the original scalar fields. This arises because of the
twisting procedure; the scalar fields transform as a vector under the original  flavor symmetry and
hence will also behave as a vector under the twisted rotational symmetry. Conversely, the original gauge
field was a singlet under the flavor symmetry so remains a vector under the twisted symmetry. Finally the
bosonic field $d$ is introduced to render the $\cQ$ symmetry nilpotent off-shell. Indeed the scalar supersymmetry
transformations take the simple form
\begin{equation}
\begin{aligned}
\cQ\, \cA_i&=\psi_i\\
\cQ\, \psi_i&=0\\
\cQ\, \cAb_i&=0\\
\cQ\, \chi_{ij}&=-\cFb_{ij}\\
\cQ\, \eta&= d\\
\cQ\, d&=0
\end{aligned}
\end{equation}
Notice that $\cQ^2=0$ on all fields as advertised. The complex $\cA_i$ yields complexified covariant
derivatives $\cD_i=\partial_i+i\cA_i$, $\cDb_i=\partial_i+i\cAb_i$ and associated field strengths
$\cF_{ij}=\left[\cD_i,\cD_j\right]$.

So far the discussion has taken place in the continuum and the entire twisting process {\it in flat space}
can be envisioned as merely an exotic change of variables. However it clearly offers some
advantages when it comes to discretization;  the twisted theory
no longer contains any spinors which makes it possible to avoid the usual fermion doubling problem.
Indeed after doing the $\cQ$-variation the fermionic part of the action describes a \KD fermion which can
be discretized without inducing fermion doubling. In fact the resultant action
can be mapped into that of (reduced) staggered fermions.
Furthermore and most importantly the scalar supersymmetry
can be restricted to a lattice without paying any penalty. 

In more detail the transcription to a lattice requires first assigning continuum fields to links in  a lattice. The
lattice is not arbitrary; in the example in question one requires a lattice with both the usual unit basis
vectors in the coordinate directions $x\to x+\hat{i}$ for $\cA_i$ and its superpartner $\psi_i$ but also diagonal
or face links running from $x+\hat{i}+\hat{j}\to x$ to carry the $\chi_{ij}$. We also need a prescription
for replacing continuum derivatives with (gauged) difference operators. Such a prescription exists
and we illustrate it below for a generic link field $f_i(x)$
\begin{eqnarray}
\cD_i f_j &\to& \cU_i(x)f_j(x+\hat{i})-f_j(x)\cU_i(x+\hat{j})\\\nonumber
\cDb_i f_i&\to& f_i(x)\cUb(x)-\cUb(x-\hat{i})f_i(x-\hat{i})
\end{eqnarray}
In this expression we have replaced the continuum $\cA_i$ by a complex lattice
Wilson link field $\cU_i$. Notice that these expressions ensure that the derivatives gauge transform
like appropriate link paths and have the correct naive continuum limit if $\cU_i=I+\cA_i+\ldots$
Furthermore, notice that this definition means that
$\cF_{ij}=\cD_i\cU_j$ is automatically antisymmetric in its indices and remarkably satisfies
an exact Bianchi identity $\epsilon_{ijkl}\cD_i \cF_{jk}=0$.

Using the exact symmetries of the lattice action allows one to strongly constrain the possible
{\it relevant} counter terms that can appear in the lattice effective action. The $\cQ$ symmetry is
particularly important in this regard. Consider the one loop effective action gotten by
expanding the gauge fields about a generic classical vacuum state $\cU_i=I+a_i+\cA_i$
where $a_i$ is a diagonal constant matrix corresponding to one of the flat directions in the theory.
It is not hard to show that the Pfaffian that results from integration over the twisted fermions
cancels a corresponding bosonic determinant and so the one loop effective action  $\Gamma(a_i)$
is zero
because of supersymmetry. However this result in fact holds to all orders; it turns out that the
vacuum expectation value of
any $\cQ$-invariant operator is independent of the coupling constant and hence can be evaluated 
exactly at one loop. The proof is straightforward. Consider
\begin{equation}
<O>=\int D\Phi\, O e^{-\beta \cQ\Psi}
\end{equation}
where we denote all fields generically by $\Phi$ and the action takes a $\cQ$-exact form as
we have described previously.
Differentiating with respect to the coupling $\beta$ yields
\begin{equation}
\frac{\partial <O>}{\partial\beta}=\int D\Phi\, O\cQ\Psi\, e^{-\beta\cQ\Psi}=\int D\Phi\, \cQ(O\Psi)\,e^{-\beta\cQ\Psi}=0\end{equation}

This result ensures that no scalar potential appears to any order in perturbation theory; the scalars
remain massless and the flat directions survive quantum correction. In addition fermion masses are also
suppressed; gauge invariance requires any operator to take the form of a closed loop. Relevant operators
correspond to loops of minimal length; most of these correspond to kinetic terms already
appearing in the classical action; the only exception being a term of the form $\cQ\Tr\,\eta\sum_i\cUb_i \cU_i$.
However this term lifts the flat directions and so is prohibited by the proceeding argument if it is not
present in the classical action \cite{Catterall:2011pd,Catterall:2014mha}

Let us wrap up this section by summarizing the key differences between this approach and more
conventional lattice formulations of (supersymmetric) gauge theories. 
\begin{itemize}
\item Fermions live in the algebra. To maintain supersymmetry so must the gauge fields. This
implies that the one must employ a flat measure rather than the Haar measure in the path integral.
This is
very different from lattice QCD. The usual problems of maintaining
gauge invariance are avoided since the links are complexified and hence the flat measure $D\cU D\cUb$ is
still gauge invariant.
\item The correct naive continuum limit requires that the $\cU_i=I+\cA_i+\ldots$. With a group valued link field
the unit matrix appearing here is automatic but when the variables reside in the algebra it needs to arise
by giving a vacuum expectation value to a dynamical field in the theory. Luckily since the gauge group is $GL(N,C)$ this
can be arranged by letting the trace mode of the (untwisted) scalar field take on such a vev. In practice
we guarantee this by adding a soft supersymmetry breaking term to the action of the form
$\sum_{x,i} \left[\frac{1}{N}\Tr\,(\cU_i(x)\cUb_i(x))-1\right]^2$
\item Supersymmetry forces fermions to be assigned to links like their superpartners the gauge  fields.
This is different from lattice QCD. In addition the fermions are treated as \KD fields. Fermion doubling is
avoided by a careful discretization procedure of the latter.
\end{itemize}

\subsubsection{Super QCD}
\label{sec:SQCD2D}
Remarkably the previous constructions can be generalized to include fermions and scalars
transforming in the fundamental representation of the gauge group. The trick is to start from
a lattice super Yang-Mills theory in one higher dimension with at least one scalar
supercharge. One then restricts the extra dimension so that it contains only two timeslices and
gauge those two timeslices under two independent gauge groups $U(C)$ and $U(F)$. 
To maintain gauge invariance the links running between these two timeslices must now contain
fields which transform in the bifundamental representation of the combined gauge group
$U(C)\times U(F)$. Furthermore the extra dimensional
gauge field will behave as a scalar with respect the twisted
rotational symmetry of each timeslice. Finally the gauge coupling for say the $U(F)$ theory
is sent to zero resulting in a theory containing both (twisted) gauge fields and fermions
in the adjoint representation of $U(C)$  (a vector multiplet) together with $F$ scalars and fermions in
the fundamental representation of that group ($F$ hypermultiplets). All fields on the $U(F)$ timeslice
can then be consistently truncated from the theory. A single exact supercharge remains and
constrains the renormalization of the lattice theory.

The key result which makes this construction possible is a generalization of the prescription
used in the case of adjoint fields to replace covariant derivatives by covariant finite difference operators to
the case of bifundamental fields. Consider a bifundamental fermion $\psi_\mu$ which one
can think of as a rectangular $C\times F$ matrix and
which transforms as
\begin{equation}
\psi_\mu(x)\to G(x)\psi_\mu H(x+\hat{\mu})\;{\rm where}\; G\in U(C)\;{\rm and}\; H\in U(F)\end{equation}
A lattice derivative can be defined as
\begin{equation}
\cD_\mu\psi_\nu(x)=\cU_\mu(x)\psi_\nu(x+\hat{\mu})-\psi_\nu(x){\cal V}_\mu(x+\hat{\nu})\end{equation}
where $\cU_\mu$ is the $U(C)$ gauge field and ${\cal V}_\mu$ the $U(F)$ gauge field. This forms
a gauge invariant loop when traced with the bifundamental $F\times C$ rectangular
fermion $\chi_{\mu\nu}(x)$.

Using these techniques one could study three dimensional super QCD but unfortunately not in four dimensions since
in the latter case one would need to start from a five dimensional theory with a single exact supercharge
which  does not exist.

\subsection{Nonlocal lattice actions}
\label{sec:noloc}
Locality is one of the basic principles that is usually required in quantum field theories. On
the other hand nonlocal lattice actions can sometimes allow us to preserve supersymmetry and circumvent the fermion doubling problem. The perturbative calculations in four dimensional gauge theories with nonlocal lattice actions  have shown that nonlocal counterterms are required to achieve a local continuum limit \cite{Karsten:1979wh}. This is one of the basic arguments why nonlocal lattice actions are considered as unusable in lattice simulations. 

On the other hand in low dimensional Wess-Zumino models the correct local continuum limit of the nonlocal lattice actions can be shown in perturbation theory \cite{Bergner:2007pu}. The perturbative proof is valid up to three dimensions \cite{Kadoh:2009sp}. 
In four dimensions nonlocal lattice representations of the Wess-Zumino model have been proposed that include supersymmetry transformations with non-linear and nonlocal modifications \cite{Feo:2013faa}. The effects of the locality violation have been investigated in numerical simulations \cite{Chen:2010uca}. Numerical evidence shows that the breaking is not severe, but no exponential localization could be observed.

The violation of locality has more severe consequences for gauge theories than for Wess-Zumino models. 
The nonlocal contribution can be introduced in terms of a sharp momentum cutoff for all fields, which leads, in case of the gauge fields, to a violation
of gauge invariance. Considering only a nonlocal fermion action to resolve the doubling problem, gauge invariance requires gauge transports at all distances. 
Hence it appears that at least for supersymmetric gauge theories and supersymmetric QCD, especially in four dimensions, this approach is not applicable.
\section{Applications and Results}
\label{sec:results}
\subsection{Wess-Zumino models and supersymmetric Yang-Mills theories in less than four dimensions}
As explained in Section \ref{sec:finetuning}  and  \ref{sec:noloc} the fine tuning problem is much simpler in lower dimensional theories and
even non-local lattice actions can be considered, at least in the case of Wess-Zumino models.

These theories have been studied as toy models to investigate supersymmetry breaking on the lattice. Wess-Zumino models in particular are interesting candidates 
for these kind of investigations. A large number of theoretical approaches have been taken to realize lattice formulations of these models. Numerical investigations
with naive and improved discretizations and also featuring partial realization of the supersymmetry have been done\cite{Catterall:2000rv,Catterall:2003ae,Bergner:2007pu}. 
Particularly interesting from the conceptual point of view are approaches based on the Nicolai map, which maps the interacting theory to a free theory\cite{Beccaria:2001tk}.
The one and two dimensional Wess-Zumino models also serve as useful arenas for investigating spontaneous supersymmetry breaking\cite{Kanamori:2007ye,Wozar:2011gu,Steinhauer:2014yaa}.

In addition to serving as toy models for understanding the general problem of lattice supersymmetry low dimensional models
can also have interesting applications in 
gauge-gravity duality -- finite temperature super Yang-Mills models in one, two and three dimensions and in the planar limit are conjectured to be dual to certain
supergravity theories containing black holes. Moving away from the planar limit allows the Yang-Mills  simulations to tell us something about quantum and string
corrections in these theories. These topics will be covered in more detail in this issue  by
Masanori Hanada \cite{Hanada:2016jok}.

\subsection{$\mathcal{N}=1$ supersymmetric Yang-Mills theory}
\label{sec:n1sym}
\begin{figure}[htb]
\centerline{\includegraphics[width=0.49\textwidth]{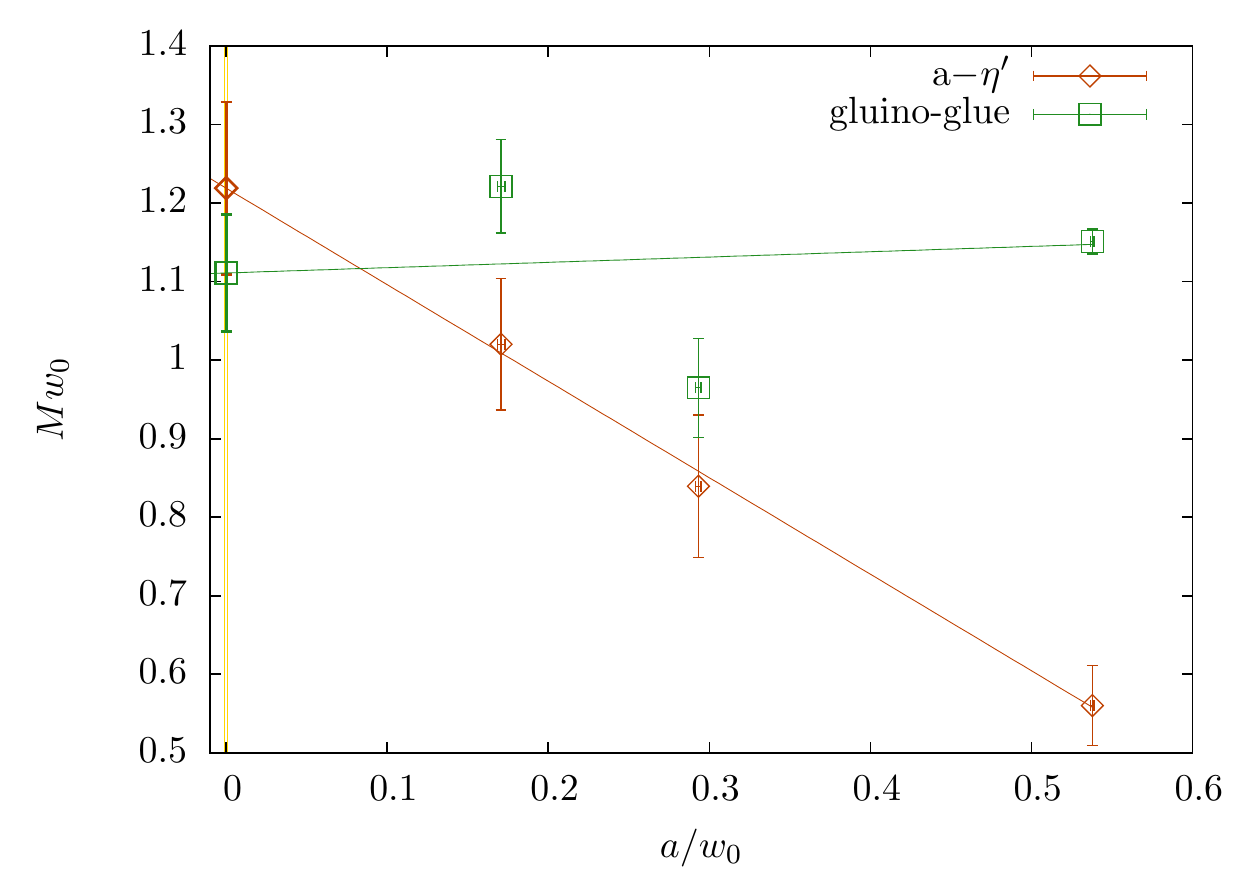}\includegraphics[width=0.49\textwidth]{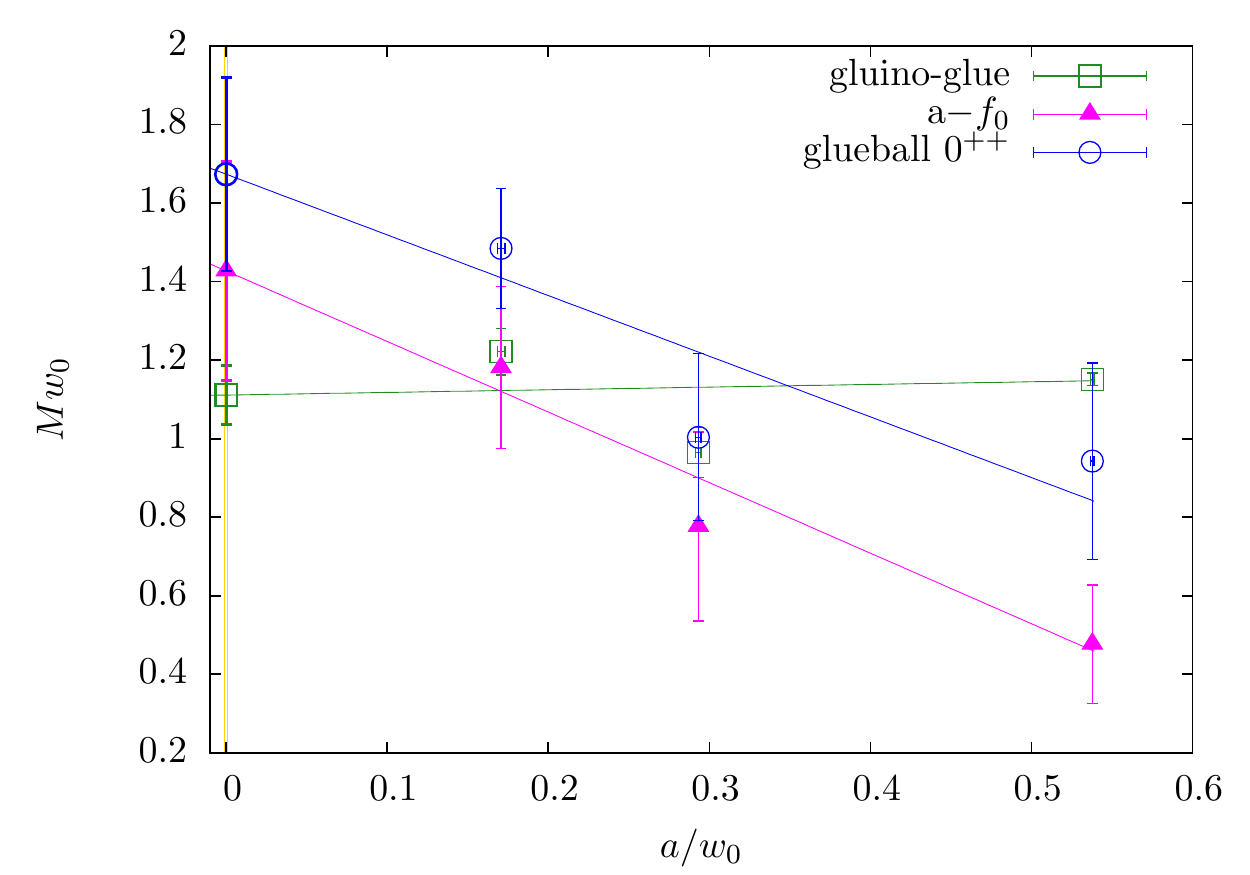}}

\caption{\label{fig:symspec} The continuum extrapolation of the lowest multiplet in $\Su(2)$ supersymmetric Yang-Mills theory\cite{Bergner:2015adz}. The multiplet consists of a scalar (represented as mesonic adjoint $f_0$ or $0^{++}$ glueball), a pseudoscalar (represented by the mesonic adjoint $\eta'$), and a fermionic particle (the gluino-glue). The particle mass $M$ and the lattice spacing $a$ are given in units of the 
scale $w_0$ determined by the Wilson flow.}
\end{figure} 
$\mathcal{N}=1$ supersymmetric Yang-Mills theory (SYM) is the pure gluonic sector of the supersymmetric extension of the standard model. It is an interesting subject
for non-perturbative investigations; not only due to its relevance in extensions of the standard model, but also because of the various theoretical considerations and 
predictions for this theory \cite{Amati:1988ft}. The theory consists of the usual gluonic Yang-Mills theory with the fermionic counterparts, the gluinos. At a first glance, the theory looks quite similar to
the massless limit of 
one-flavor QCD albeit with Majorana fermions in the adjoint representation. Besides supersymmetry, the theory has an $\mathrm{U}(1)$ R-symmetry, which corresponds the 
chiral symmetry group of a theory with one fermion flavor. The anomaly breaks this symmetry to a discrete $Z_{2N_c}$ subgroup in the case of an $\Su(N_c)$ gauge group.
This remaining symmetry is broken by a fermion condensate down to $Z_2$.

At low energies the gluinos and gluons are confined in strongly bound states. If supersymmetry is unbroken, the bound state spectrum should be composed of supersymmetry multiplets.
These multiples consist of a bosonic scalar, a bosonic pseudoscalar, and a fermionic particle with the same mass. Low energy effective theories have been constructed with 
the first multiplet of gluonic type, where the bosonic operators are glueballs, and a second one of mesonic type, with meson like gluino-ball operators as bosonic constituents \cite{Veneziano:1982ah,Farrar:1997fn,Farrar:1998rm}.
Further interesting predictions for ${\cal N}=1$ SYM are the exact value of the gluino condensate and the all order beta function \cite{Shifman:1987ia,Novikov:1983uc}.

As explained in Section \ref{sec:finetuning}, the restoration of supersymmetry in the continuum limit can be achieved relatively easily in this theory.
Using Ginsparg-Wilson fermions, it is obtained without fine tuning. For Wilson fermions  a single fine tuning of the fermion mass 
is sufficient. In practice, the best signal for the fine tuning is the breaking of chiral symmetry in terms of the adjoint pion mass. 
This particle is defined in partially quenched chiral perturbation theory \cite{Munster:2014cja}.

In addition to these theoretical considerations one has to face several technical challenges in the simulations of ${\cal N}=1$ SYM.  The theory contains a Majorana
fermion (in the adjoint representation) which yields a 
Pfaffian after integrating out the fermions unlike the usual determinant encountered in lattice QCD. Even if the determinant can be proven positive this is
not necessarily true for the Pfaffian and so this theory suffers from a sign problem. In practice this is not too severe and can be handled
using reweighting techniques. In addition, the bound states of the theory are either gluonic observables or flavor singlet mesonic states, both of them are rather hard to measure. 

Compared to the determination of the bound state spectrum, the measurement of the gluino condensate can be performed more easily. In the first investigations 
using Wilson fermions the chiral phase transition was
determined from the two peak structure of the histogram of the condensate \cite{Kirchner:1998mp}. However, with Wilson fermions this quantity includes additive and 
multiplicative renormalization and therefore investigations with Ginsparg-Wilson fermions are favored. Interesting results have been obtained 
using simulations with domain wall fermions
\cite{Giedt:2008xm,Endres:2009yp}.

The determination of the bound state spectrum is a more challenging task and it has so far only been done with Wilson fermions. 
While the first preliminary investigations can be found in \cite{Montvay:1995ea} some more recent results are presented in \cite{Demmouche:2010sf}. These
latter results profit from improved dynamical fermion algorithms and incorporate extrapolations to the chiral limit, but they have found a rather large 
splitting between the bosonic and fermionic components of the lowest multiplet.
 A careful analysis of the lattice artifacts and the finite size effects was necessary 
to resolve this issue \cite{Bergner:2012rv,Bergner:2013nwa}. The final results are consistent with the formation of a multiplet of bound states \cite{Bergner:2015adz}, see Figure \ref{fig:symspec}.

The pure ${\cal N}=1$ SYM has a number of interesting applications and the lattice simulations might confirm the theoretical conjectures about this theory. The first lattice simulations of 
$\Su(2)$ ${\cal N}=1$ SYM at finite temperature have found a second order deconfinement transition at around $0.8 (T_c)_{\mathrm{YM}}$ compared to the critical temperature in pure $\Su(2)$ 
Yang Mills theory $(T_c)_{\mathrm{YM}}$. The chiral phase transition happens at around the same temperature\cite{Bergner:2014saa}. 

\begin{figure}[htb]
\centerline{\includegraphics[width=\imscale\textwidth]{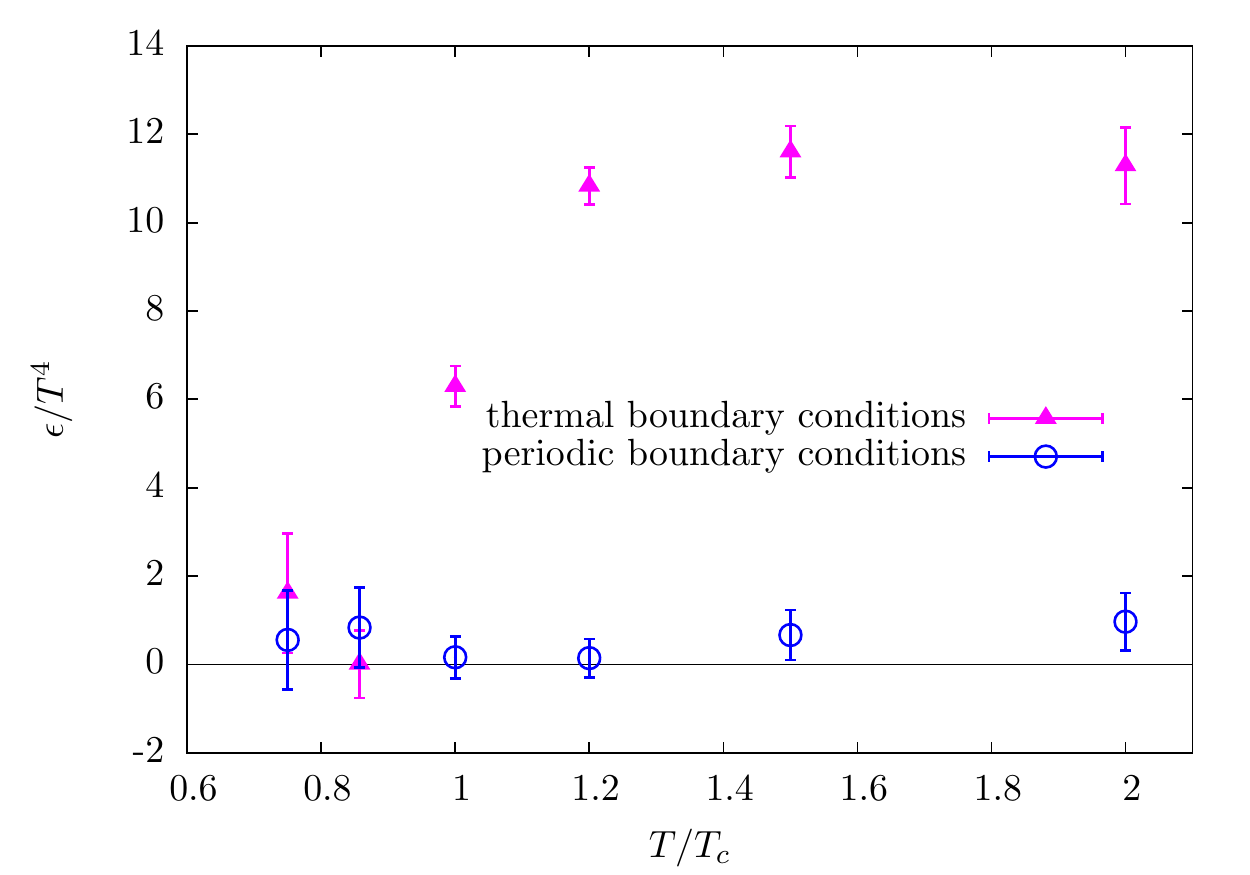}}
\caption{\label{fig:susycompact} This Figure shows a comparison of the derivative of the thermal partition function and the twisted partition function for $\Su(2)$ ${\cal N}=1$ SYM.
In contrast to the thermal case, periodic boundary conditions are applied for the fermions in the twisted partition function\cite{Bergner:2015cqa}. The temperature is identified with $1/R$, where 
$R$ is the compactification radius of the compactified theory on ${R^3\times S^1}$. The simulations have been done with a tree level clover improved  fermion action and 
an adjoint pion mass of around $am_\pi\simeq 0.6$.}
\end{figure}

Recent theoretical investigations consider compactified ${\cal N}=1$ SYM on ${R^3\times S^1}$. Instead of the thermal boundary conditions, which are antiperiodic for the fermions and periodic for the bosons,
 periodic boundary conditions are applied in the compactified direction. In this compactified theory no phase transition is expected even down to a small radius, where the theory
can be understood by means of a semiclassical analysis\cite{Poppitz:2012sw}. Lattice simulations were able to identify indications of the 
expected continuity\cite{Bergner:2014dua,Bergner:2015cqa}. One interesting example is shown in Figure \ref{fig:susycompact}. With thermal boundary conditions the quantity $\epsilon$ corresponds to the volume averaged 
derivative of the partition function with respect to the temperature and provides information about the equation of state. With periodic boundary conditions, it is a derivative of the twisted partition function and corresponds to a sum of differences between bosonic and fermionic energy levels. Hence this measurement provides an indication about the absence of the phase transition with periodic boundary conditions and about the smallness of the 
remnant supersymmetry breaking on the lattice.

\subsection{$\mathcal{N}=4$ supersymmetric Yang-Mills theory}
The twisting procedure described in Section \ref{sec:plsusy} can be applied to $\cN=4$ Yang-Mills and results in
a lattice theory which retains one exact supercharge at non zero lattice spacing. The action for this lattice theory
is very similar to that given in the two dimensional example - see for example the review \cite{Catterall:2009it}
\begin{equation}
S=\cQ \sum_x \Tr\, \left(\sum_{a,b=1}^5\chi_{ab}\cF_{ab}+\eta \sum_{a=1}^5\cDb_a\cU_a+\frac{1}{2}\eta d\right)+S_{\rm closed}\end{equation}
Notice that the ten bosonic fields of $\cN=4$ (4 gauge fields and six scalars) are packed into
5 complex gauge fields $\cU_a,a=1\ldots 5$
while the sixteen fermionic degrees of freedom appear as $\left(\eta,\psi_a,\chi_{ab}\right)$.
The term $S_{\rm closed}$ is a new term that only appears in four dimensions. It takes the form
\begin{equation}
S_{\rm closed}=\sum_x \epsilon_{abcde}\chi_{ab}\cDb_c\chi_{de}
\end{equation}
This term is invariant under $\cQ$ by virtue of the exact lattice Bianchi identity (recall that
$\cQ\,\chi=-\cFb$)
In \cite{Catterall:2014mha} we showed that this lattice theory
requires at most a single tuning of a marginal coupling to
target the continuum $\cN=4$ theory in the continuum limit in which all the supersymmetries are
restored \cite{Catterall:2013roa}. One should place a caveat on this
result; the relevance of any operator depends on a power counting argument using the
engineering dimension of a field - it is possible that at strong coupling  large anomalous dimensions
can be generated and modify the set of relevant operators.
\begin{figure}[htb]
\centerline{\includegraphics[width=\imscale\textwidth]{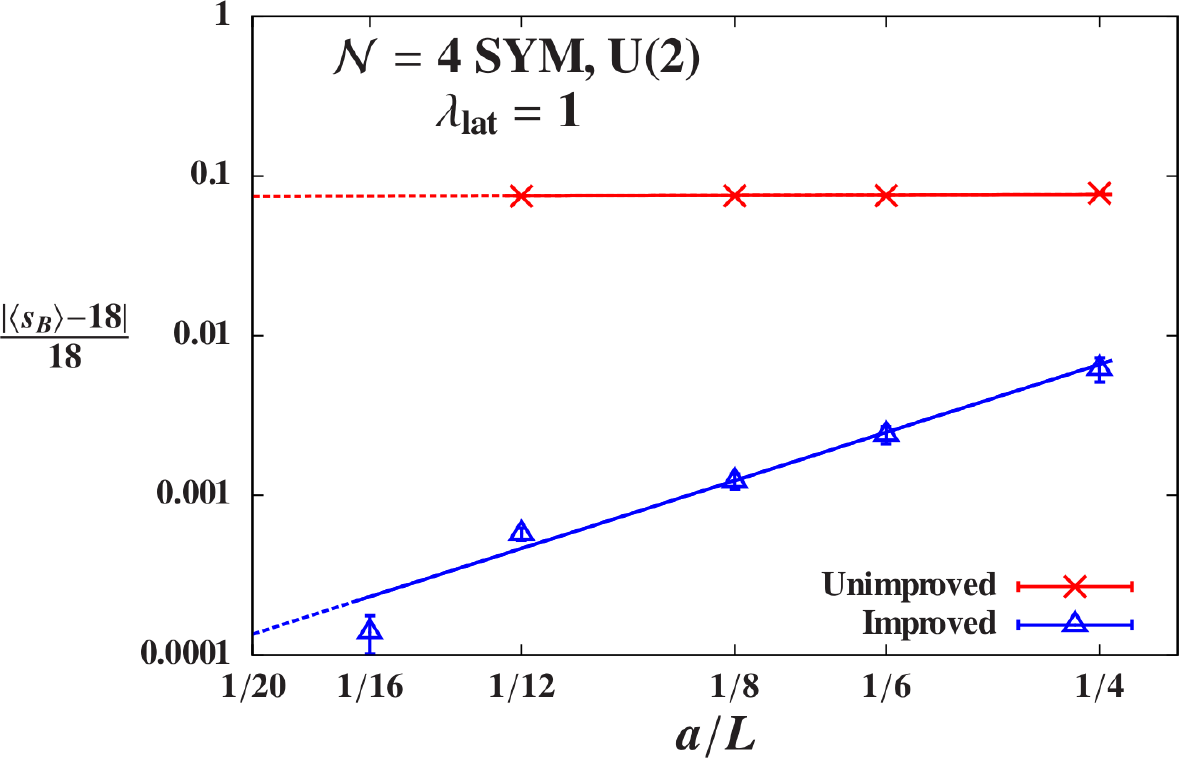}}
\caption{\label{susyward} The Ward identity in $\Su(2)$ $\mathcal{N}=4$ SYM as a function of the inverse lattice size at fixed $\lambda=1$ is compared for an improved and unimproved lattice formulation.}
\end{figure}
As we have discussed it is necessary to add a soft $\cQ$-breaking potential to pick out the
vacuum state $\cU_a=I+\ldots$ by adding a potential of the form
\begin{equation}
\delta S=\mu^2\sum_{x,i} \left[\frac{1}{N}\Tr\,(\cU_i(x)\cUb_i(x))-1\right]^2\end{equation}
In practice the $\mu^2$ dependence is rather weak and yields rather robust
extrapolations $\mu^2\to0 $. 

However in four dimensions we observe a second problem; at strong coupling
we observe a condensation of lattice monopoles associated with the $U(1)$ gauge field. To remove
this lattice artifact we have modified the action by the addition of another $\cQ$-exact term with
coupling $G$
\begin{equation}
G\,\cQ \sum_{x,a,b} \Tr\, \left[\eta \left({\rm det} P_{ab}-1\right)\right]\end{equation}
where $P_{ab}$ is the (complexified) Wilson plaquette operator.
This changes the moduli space of the theory to include  only $SL(N,C)$ configurations and yields a new potential
term of the form $\left[\left({\rm det} P_{ab}-1\right)\right]^2$ which penalizes fluctuations of the determinant
of the
plaquette away from unity. Since this modification of the action is supersymmetric the violations of
$\cQ$ Ward identities are small as can be seen in Figure \ref{susyward} which compares the
improved action over an earlier iteration where the monopoles are suppressed in a way
which breaks supersymmetry.

\begin{figure}[htb]
\centerline{\includegraphics[width=\imscale\textwidth]{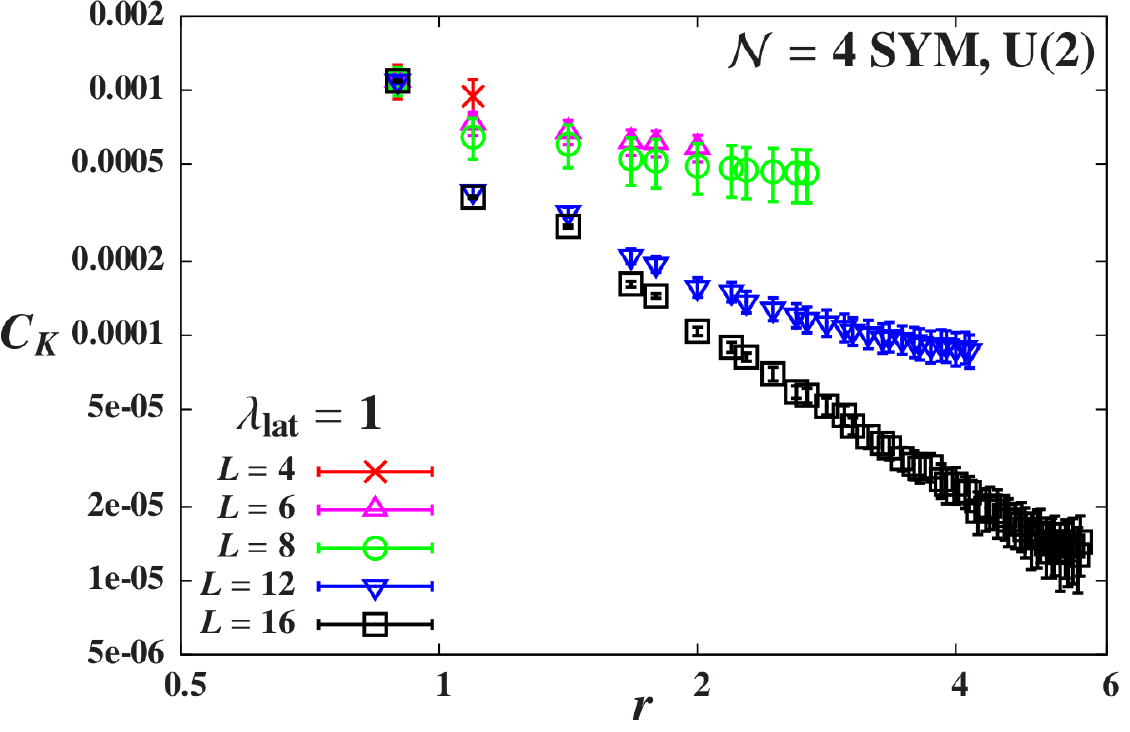}}
\caption{\label{correlators} The two point function of the Konishi operator at $\lambda=1$ in $\Su(2)$ $\mathcal{N}=4$ SYM.}
\end{figure}

Using a parallelized code based on the MILC libraries \cite{Schaich:2014pda} we are currently using lattice simulation
to probe the structure of $\cN=4$ Yang-Mills at strong coupling and for small numbers of colors. This
is a regime inaccessible to analytic computations which typically require taking the planar limit. It
allows us to search for signs of S-duality \cite{Montonen:1977sn,Kapustin:2006pk} and to test the bounds on anomalous dimensions provided
by the conformal bootstrap program \cite{Beem:2013qxa}. 
\begin{figure}[htb]
\centerline{\includegraphics[width=\imscale\textwidth]{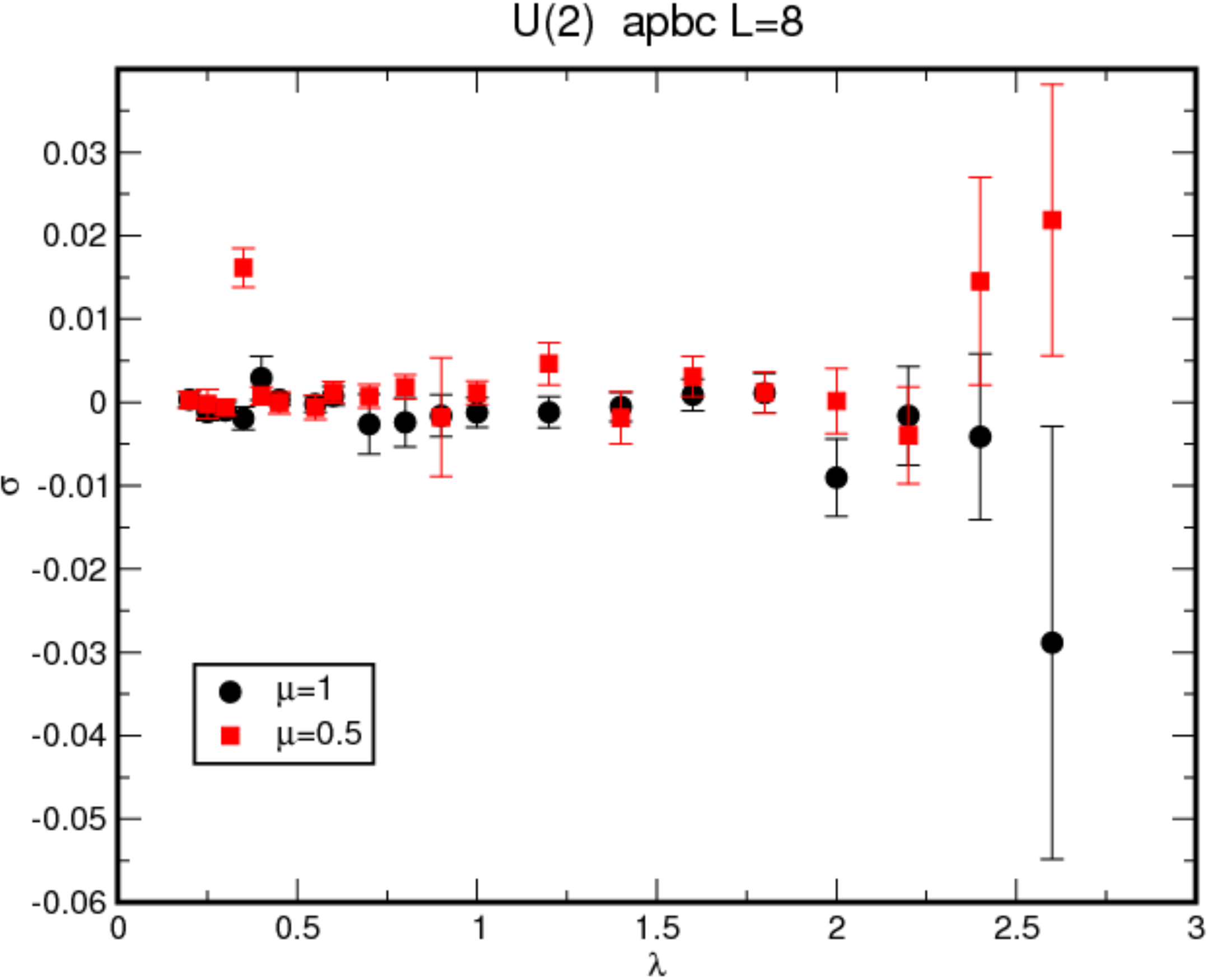}}
\caption{\label{sigma} The string tension as a function of the 't Hooft coupling in $\Su(2)$ $\mathcal{N}=4$ SYM.}
\end{figure}
One of the central features of $\cN=4$ Yang-Mills that we would
like to reproduce is the fact that it is conformal for any value of the
gauge coupling. To this end we have computed the static potential $V(r)$
from the correlators of Wilson lines
(after gauge fixing). We find that fits to the form $V(r)=\sigma r+\frac{C}{r}$ always yield a string
tension $\sigma\sim 0$ within errors - see Figure \ref{sigma}. This is consistent with the system
being in a conformal phase for all couplings. Furthermore, the Coulomb coefficient $C$ is found to
agree with perturbative estimates at weak coupling.
Further evidence in favor of conformality can be found by examining the behavior of two point
functions of would be conformal operators in the theory. The simplest of these is the Konishi
operator - the flavor singlet quadratic scalar operator given by
\begin{equation}
O_K=\sum_I \Tr\,\left(\phi^I\phi^I\right)=\sum_a \frac{1}{N}\Tr\,\left(\cU_a\cUb_a\right)-1\end{equation}
This is shown in Figure \ref{correlators} in a log-log plot. The increasing
linearity of the plot as the lattice size increases is very consistent with a power law behavior in the
infinite volume limit.
In principle the slope of this line yields (twice) the scaling dimension of the Konishi operator and
we are currently working hard to extract this scaling dimension as a function of the 't Hooft coupling to
compare with results in the planar limit \cite{Gromov:2009tv} and bounds from the conformal bootstrap approach \cite{Beem:2013qxa}.
Preliminary results obtained from a Monte Carlo renormalization group analysis are in agreement
with perturbative calculations at weak coupling.

\subsection{Towards supersymmetric QCD}
$\mathcal{N}=1$ supersymmetric QCD is obtained when the supersymmetric pure gauge theory presented in Section \ref{sec:n1sym}
is coupled to a Wess-Zumino model with fields in the fundamental representation. The complete solution for the correct representation 
of the four dimensional Wess-Zumino model on the lattice without fine tuning are so far unknown, but first results indicate, that
constructions guided by perturbative arguments might offer a reasonable solution\cite{Chen:2010uca}. 
Supersymmetric QCD in four dimension requires a large number of terms to be fine tuned in the continuum limit and the practical 
applicability of the fine tuning program is so far unknown. Similar considerations also hold for $\mathcal{N}=2$ SYM in four dimensions.

\begin{figure}[htb]
\centerline{\includegraphics[width=\imscale\textwidth]{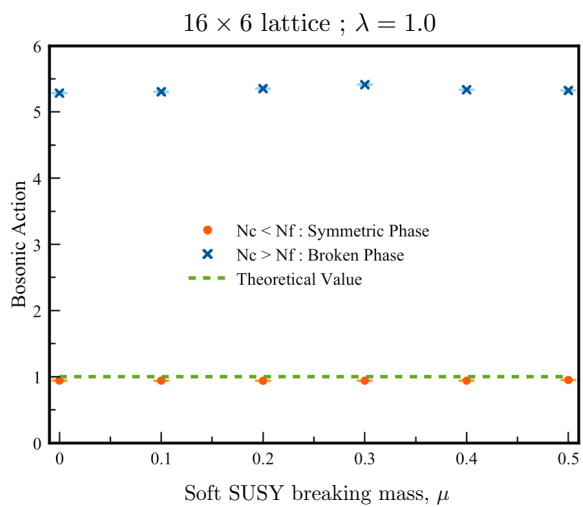}}
\caption{\label{Bact} The Bosonic action as a function of the soft supersymmetry breaking mass parameter $\mu$ in two dimensional supersymmetric QCD.}
\end{figure}
It is therefore instructive to study first an example in two dimensions, where as discussed in Section \ref{sec:SQCD2D} a lattice 
discretization with the correct continuum limit can be found.
This model was studied with the addition of a $\cQ$-invariant
Fayet-Iliopoulos term $r\cQ\sum_x\Tr\, \eta(x)$\cite{Catterall:2015tta}. The addition of this term allows for dynamical
supersymmetry breaking -- after integrating out the auxiliary $d$-field on the $U(C)$ lattice
one finds a potential term of the form
\begin{equation}
\delta V=\sum_x \Tr\,\left(\sum_{f=1}^F\phi^f\overline{\phi}^f-rI_{C}\right)^2
\end{equation}
where $\phi$ is the bifundamental scalar resulting from dimensional reduction in the extra dimension
and the trace runs over $C$ colors. Whether one can set this potential to zero (and hence
find a supersymmetric vacuum) depends on the rank of the $C\times C$ matrix $\phi\overline{\phi}$.
One expects for $F\ge C$ that a supersymmetric vacuum is possible while for $F<C$ dynamical
supersymmetry breaking should occur. The numerical results in \cite{Catterall:2015tta} bear this out and also
find evidence for a Goldstino in the latter case. Figure \ref{Bact} plots the bosonic action which can be
obtained via a $\cQ$ Ward identity in both cases. The dashed line indicates the result expected
for a theory in which supersymmetry is {\it not} broken. Clearly the numerical results are completely
in agreement with the theoretical arguments based on the rank of scalar potential.

\section{Conclusions}
In this brief review we have listed some of the problems faced when studying supersymmetric theories on lattices. In general
supersymmetry will be broken completely and one faces a fine tuning problem to regain a supersymmetric theory in the limit in
which the lattice spacing is sent to zero. In low dimensions this can sometimes be avoided  either by using non-local actions or
by performing a finite order perturbative calculation to determine the coefficients of the counter terms. In certain cases this
fine tuning problem can be reduced or even eliminated using new lattice actions which conserve (at least) a single supercharge.
This latter situation includes ${\cal N}=4$ super Yang-Mills and its dimensional reductions. These supersymmetric
actions are discussed in some detail in the review. For ${\cal N}=1$ super Yang-Mills theory
the fine tuning problem involves only a single coupling where it coincides with the usual tuning needed to take the chiral
limit. We present encouraging new results from the numerical simulations of these theories, in particular concerning the
mass spectrum of ${\cal N}=1$ super Yang-Mills. They indicate that the theoretical and numerical challenges are now under control 
and the lattice can be an interesting tool for further non-perturbative investigations of these theories. As an example of how theoretical 
conjectures can be tested, we have shown the phase transitions in compactified ${\cal N}=1$ super Yang-Mills theory and the indications for
a conformal behavior in ${\cal N}=4$ super Yang-Mills.

While theories like super QCD remain a goal of this program, current work on these more general theories is limited to
low dimensions where encouraging results have been obtained on models which exhibit dynamical supersymmetry
breaking.

\bibliography{biblio}
\end{document}